\documentstyle[sprocl]{article}

\begin{document}

\hfill\vbox{
  \hbox{OHSTPY-HEP-T-97-011}
  \hbox{hep-ph/9707267}
}\par
\bigskip

\title{ON PERTURBATIVE QCD AT FINITE TEMPERATURE}

\author{ A. NIETO }

\address{
  The Ohio State University, Department of Physics, 
  Columbus, OH 43210~\footnote{ Address after September 1,
  1997: CERN -- Theory Division, CH-1211 Geneva 23,
  Switzerland.}
}

\maketitle\abstracts{ 
Effective field theory methods provide a convenient approach
to study static observables in field theory at finite
temperature. In this talk, I will outline the construction
of the effective field theory that describes effective
observables in QCD at high temperature. An analysis of the
convergence of the perturbative series for the free energy
of QCD will also be presented.
}

\section{Introduction}

There is a number of interesting scenarios which are
described by quantum field theories at finite temperature
(electroweak phase transition, heavy ion collisions, etc.)
In this talk I will focus on the deconfinement phase
transition of QCD: lattice simulations suggest that hadron
matter undergoes a transition into a phase in which quarks
are deconfined at a temperature of order 150 MeV. This new
state of matter called {\em quark-gluon plasma (QGP)} is
expected to be studied experimentally at CERN and Brookhaven
in the next decade. Here, we will consider the system at
high temperature so that it is well inside the QGP
phase. The temperature is the typical energy of the system
and, when it is high enough, the strong coupling constant
is small so that one would expect perturbation theory
to describe the QGP accurately.

It is specially interesting to consider the limit in which
all of the masses scales of the system are smaller than the
temperature. However, in this case, particles are massless
and one has to deal with infrared divergences. In
perturbation theory the infrared problems are fixed by
resumming infinite series of diagrams that provide
infrared-finite results.

The fields that describe a system of particles at finite
temperature in the imaginary time formalism have to satisfy
periodic (for bosons) or antiperiodic (for fermions)
boundary conditions in the Euclidean time variable. This
allows us to expand a generic field $\Phi({\rm x},\tau)$ in
terms of its Fourier modes $\phi_n({\rm x})$:
\begin{equation}
  \Phi({\rm x},\tau) = T \sum_n \phi_n({\rm x}) e^{-i\omega_n\tau} \,.
\end{equation}
In this expression, $\omega_n$ is the so-called {\em Matsubara
frequency\/} and it is $2n\pi T$ for bosons and $(2n+1)\pi
T$ for fermions where $n$ is an integer. We can
associate a propagator
\begin{equation}
  {1\over \omega_n^2 + p^2}
\end{equation}
to the $n$th-mode and $\omega_n$ then plays the role of its
mass. When $\omega_n=0$ the Fourier mode is massless and is
called static.  Otherwise, the Fourier mode has a mass
$\omega_n\sim T$ and it is called non-static.

If we decouple the non-static modes what remains is a
3-dimensional effective field theory of the static
modes. This procedure is known as {\em dimensional
reduction\/}~\cite{appelquist-pisarski,nadkarni-1,gpy}. The
effective theory is therefore described by an effective
Lagrangian ${\cal L}_{eff}$ which has to be compatible with
the symmetries of the original theory. It can be written as
${\cal L}_{eff} = {\cal L}_R + \delta{\cal L}$, where ${\cal
L}_R$ represents the renormalizable terms of the effective
Lagrangian and $\delta{\cal L}$ represents an infinite
series of non-renormalizable terms. The effective coupling
constants are determined by matching physical quantities in
the full theory and in the effective theories. It is
convenient to introduce an infrared cutoff in both theories
before matching. The cutoff is chosen to satisfy
$gT\ll\Lambda_{IR}\ll T$ so that the strict perturbation
expansion (without resummations) is enough to determine the
parameters of the effective theory by matching with the full
theory~\cite{bn5,nieto}.

\section{QCD}

In this section we will apply these ideas to compute the
free energy of QCD. Our goal is to compute the free energy
of QCD up to order $g^5$.  Therefore, we will systematically
ignore higher order contributions.

The thermodynamic properties of QCD matter in thermal
equilibrium at temperature $T$ are described by the free
energy $F_{\rm QCD}$ which is given by the expression
\begin{equation}
  e^{-\beta F_{\rm QCD}} = {\cal Z}_{\rm QCD} =
    \int {\cal D}A_{\mu}({\bf x},\tau) \,
         {\cal D}q \, {\cal D}\overline{q} \;
    e^{
      -\int_0^{\beta} d\tau \int d^3 x \, {\cal L}_{\rm QCD}
    } \,,
\end{equation}
where ${\cal Z}_{\rm QCD}$ is the partition function,
$\beta=1/T$, and ${\cal L}_{\rm QCD}$ is the Lagrangian of
QCD
\begin{equation}
  {\cal L}_{\rm QCD} = 
    {1\over 4} (G_{\mu\nu}^{a})^2 \, + \,
    \overline{q}\gamma_{\mu}D_{\mu}q \,,
\end{equation}
where $G_{\mu\nu}^{a}$ is the strength tensor and $D_{\mu}$
the covariant derivative. The gauge coupling constant is
$g$. Note that the parameters of this theory are $T$ and $g$
and the free energy {\em density\/} is given by
\begin{equation}
  {\cal F} = - {T\over V} \log{ {\cal Z}_{\rm QCD}} \,.
\end{equation}

By dimensional reduction at high $T$ we obtain the effective
theory described by the partition function
\begin{equation}
  {\cal Z}_{\rm EQCD} =
    \int  {\cal D}A_{0}({\bf x}) \, {\cal D}A_{i}({\bf x}) \;
    e^{
      -\int d^3 x \, {\cal L}_{\rm EQCD}
    } \,.
\end{equation}
Here, EQCD stands for {\em Electrostatic QCD\/}. The
Lagrangian is
\begin{equation}
  {\cal L}_{\rm EQCD} = {1\over 4} (G_{ij}^{a})^2 \,+\,
    {1\over 2} (D_i A_0^a) (D_i A_0^a) \,+\,
    {1\over 2} m_E^2 A_0^a A_0^a \,+\,
    {1\over 8}\lambda_E (A_0^a A_0^a)^2 \,+\,
    \delta{\cal L}_{\rm EQCD} \,,
\end{equation}
where, $\delta{\cal L}_{\rm EQCD}$ represents an infinite
series of non-renormalizable terms. The effective parameters
are the effective gauge coupling constant $g_E$, the
effective mass of the electrostatic field $m_E$,
$\lambda_E$, and an infinite number of coupling constants
associated with the non-renormalizable terms. In addition,
since we intend to compute the free energy, we also have to
consider the effective parameter, $f_E$, associated with the
unit operator. All of these effective parameters are
functions of $g$ and $T$.  $\lambda_E$ as well as the
parameters of the terms in $\delta{\cal L}_{\rm EQCD}$ can
be ignored since they only contribute to the free energy at
order larger than $g^5$~\cite{bn4,bn5,nieto}. Note that the
quarks have been integrated out because their Matsubara
modes have a mass of order $T$. The free energy of QCD can
be written as
\begin{equation}
  {\cal F} = T \left( f_E - {\log{{\cal Z}_{EQCD}}\over V} \right) \,.
\end{equation}

The parameters for EQCD are determined by matching the
strict perturbation expansions for the free energy and the
electric screening mass in QCD and
EQCD~\cite{bn5,nieto}. They can be written as an expansion
in powers of $g^2$.
\begin{eqnarray}
  f_E   & = & T^3       \, (\mbox{\rm Series in $g^2$}) \,, \\
  m_E^2 & = & g^2 T^2   \, (\mbox{\rm Series in $g^2$}) \,, \\
  g_E^2 & = & g^2 T     \, (\mbox{\rm Series in $g^2$}) \,, 
\end{eqnarray}

The mass of the electrostatic field $A_0^a({\rm
x})$ is $m_E\sim gT$. Therefore, for processes
involving scales much smaller than $gT$ (e.g., of order $g^2
T$), $A_0^a({\rm x})$ decouples. What remains is an
effective field theory in 3 dimensions which only depends on
the magnetostatic fields $A_i^a({\rm x})$ and it is
therefore called {\em Magnetostatic QCD\/}.  It is described
by the partition function
\begin{equation}
  {\cal Z}_{MQCD} =
    \int  {\cal D}A_{i}({\bf x}) \;
    e^{
      -\int d^3 x \, {\cal L}_{\rm MQCD}
    } \,.
\end{equation}
where 
\begin{equation}
  {\cal L}_{MQCD} = {1\over 4} (G_{ij}^a)^2 + \delta{\cal L}_{MQCD}\,.
\end{equation}
In this expression, the strength tensor now involves the
effective coupling constant $g_M$ and $\delta{\cal
L}_{MQCD}$ contains the infinite series of
non-renormalizable terms. In addition, we also have to
consider the coefficient $f_M$ of the unit operator. The
parameters of MQCD are functions of the parameters of
EQCD. The free energy of QCD is now written
\begin{equation}\label{free}
  {\cal F} = T \left( f_E + f_M - {\log{{\cal Z}_{MQCD}}\over V} \right)\,.
\end{equation}

$f_M$ is determined by matching the
strict perturbation expansions for the free energy in EQCD and
MQCD~\cite{bn5,nieto}. It can be written as an expansion
in powers of $g$.
\begin{equation}
  f_M = m_E^3 \, (\mbox{\rm Series in $g$}) \,, \\
\end{equation}
At lowest order in $g$, it can also be
shown~\cite{bn5,nieto} that
\begin{equation}
  {\log{{\cal Z}_{MQCD}}\over V} \sim g^6 T^3 \, .
\end{equation}
Therefore, this term does not contribute to the calculation
of the free energy of QCD at order $g^5$. 

We conclude that the free energy of QCD up to order $g^5$ is
given by
\begin{equation}
  {\cal F} = T \left( f_E(g,T) + f_M(m_E,g_E) \right)\,.
\end{equation}
Details of the calculations of $f_E$, $m_E$, $g_E$, and
$f_M$ and complete results are given in Refs.~4 and~5.  The
result found in this way for ${\cal F}$ is in complete
agreement with the one obtained by Kastening and
Zhai~\cite{kastening-zhai}.

\section{Convergence of Perturbation Theory}

We have outlined the calculation of the free energy as a
perturbation expansion in powers of $g$ up to order $g^5$.
In this section, we examine the convergence of the series
and it is based on Ref.~5. We now ask how small
$\alpha_s\equiv g^2/(4\pi)$ must be in order for the
perturbation expansion to be well-behaved in the sense that
every term of a given order is smaller than all of the terms
of lower order. If the series is apparently convergent in
this sense, then it can plausibly be used to evaluate the
free energy.

We study the expansion the of the free energy of QCD in
powers of $\sqrt{\alpha_s}$ at different temperatures
$T>T_c\sim 200$ MeV. We choose the renormalization scale to
be $\mu = 2 \pi T$, which is the mass of the lightest
non-static mode. In Table~\ref{table1} we give the
contributions coming from each order in $\sqrt{\alpha_s}$,
rescaled so that the leading order term is 1, the next terms
are of order $\alpha_s$, $\alpha_s^{3/2}$, $\alpha_s^4$, and
$\alpha_s^{5/2}$, respectively. We see that the correction of order
$\alpha_s^{5/2}$ is the largest unless the temperature $T$
is greater than 2 GeV. Also, the term of order $\alpha_s^{3/2}$ is
smaller than the term of order $\alpha_s$ only when the
temperature is greater than about 1 TeV.
\begin{table}
  \begin{center}
  \begin{tabular}{ccc}
    $T$ (GeV)   & $\alpha_s(2\pi T)$    & expansion for $F$  \\ \hline
    0.250       & 0.321                 & $1-0.282+0.583+0.276-1.094$ \\
    0.500       & 0.239                 & $1-0.210+0.374+0.010-0.524$ \\
    1           & 0.194                 & $1-0.167+0.267+0.033-0.287$ \\
    2           & 0.165                 & $1-0.142+0.209+0.011-0.191$ \\
    1000        & 0.074                 & $1-0.062+0.062+0.011-0.024$
  \end{tabular}
  \caption{Perturbation expansion for the free energy of QCD
  at different temperatures.}
  \label{table1}
  \end{center}
\end{table}

We can go further in our analysis provided that we have
separated the contributions from the scales $T$ and $gT$.
The term $f_E$ gives the contribution to the free energy
from the scale $T$ and $f_M$ gives the contribution from the
scale $gT$. The analysis shows that the perturbation series
for $f_E$ is well-behaved even at temperatures as low as 250
MeV. On the contrary, the temperature for which the
perturbation series of $f_M$ is well-behaved is larger that
2 GeV.  This suggests that the slow convergence of the
expansion for the free energy of QCD in powers of
$\sqrt{\alpha_s}$ may be attributed to the slow convergence
of perturbation theory at the scale $gT$.

\section*{Acknowledgments}
This work was supported in part by the U.S. Department of
Energy, Division of High Energy Physics, under Grant
DE-FG02-91-ER40690.

\end{document}